\documentclass[12pt]{iopart}
\usepackage{iopams}
\usepackage{amstext}
\usepackage{setstack}
\usepackage{graphicx}
\usepackage{dcolumn}
\usepackage{bm}
\usepackage{hyperref} 
\usepackage{braket}
\usepackage{color}
\usepackage{framed}
\DeclareGraphicsExtensions{.pdf,.png}
\usepackage[bottom = 3 cm]{geometry}
\usepackage{multirow}
\usepackage{epstopdf}

\newcommand{\figref}[1]{Fig.~\ref{#1}}
\newcommand{\tabref}[1]{Table~\ref{#1}}

\begin{document}
\title{Optimal randomness generation from optical Bell experiments}

\author{Alejandro M\'{a}ttar$^1$, Paul Skrzypczyk$^1$, Jonatan Bohr Brask$^2$, Daniel Cavalcanti$^1$ and Antonio Ac\'{i}n$^{1,3}$}

\address{$^1$ ICFO-Institut de Ci\`encies Fot\`oniques, Mediterranean Technology Park, 08860 Castelldefels (Barcelona), Spain}
\address{$^2$ Department of Theoretical Physics, University of Geneva, 1211 Geneva, Switzerland}
\address{$^3$ ICREA-Instituci\'o Catalana de Recerca i Estudis Avan\c cats, Lluis Companys 23, 08010 Barcelona, Spain}
\ead{alejandro.mattar@icfo.es}

\begin{abstract}
Genuine randomness can be certified from Bell tests without any detailed assumptions on the working of the devices with which the test is implemented. An important class of experiments for implementing such tests is optical setups based on polarisation measurements of entangled photons distributed from a spontaneous parametric down conversion source. Here we compute the maximal amount of randomness which can be certified in such setups under realistic conditions. We provide relevant yet unexpected numerical values for the physical parameters and achieve four times more randomness than previous methods.
\end{abstract}
\section{Introduction}
\label{sec.intro}
Quantum systems have the potential to provide a strong form of randomness which cannot be attributed to incomplete knowledge of any classical variable of the system. At the basis of such genuine randomness lies a quantitative relation between the amount by which a Bell inequality is violated \cite{bell1964} and the degree of predictability of the results of the test \cite{Pironio2010}. Intuitively, the violation of a Bell inequality  certifies the presence of nonlocal correlations \cite{BellReview}, and in turn, this guarantees that the outcomes of the measurements cannot be determined in advance \cite{Ekert91,Colbeck07}. Furthermore, this genuine randomness can be certified without any detailed assumptions about the internal working of the devices used, that is, in a ``device-independent'' fashion. Device independence is advantageous since it provides immunity to attacks that exploit imperfections in the physical implementation, to which device-dependent protocols are susceptible \cite{Gerhardt2011}. For this reason, device-independent randomness generation has recently received much attention \cite{Pironio2014,Bancal2014,Bancal2014b,Bancal2014c,Dhara2014,DelaTorre2014}.

An intense research effort has been devoted to the experimental realisation of device-independent randomness generation. A few years ago, Pironio \textit{et al.} \cite{Pironio2010} implemented the first proof-of-principle experiment. It involved two entangled atomic ion qubits confined in two independent vacuum chambers separated by approximately one meter. This implementation, which was based on light-matter interaction, managed to certify 42 random bits over a period of one month. 

The principal challenge for a device-independent randomness generation experiment is that it must close the detection loophole \cite{Pearle1970,Santos92}, \textit{i.e.} it must provide a Bell inequality violation without post-selection on the data, since otherwise violation can be faked by classical resources \cite{Gerhardt2011b} and no genuine randomness can be guaranteed. The detection loophole was first successfully closed on several systems relying on light-matter interaction; see for instance \cite{Rowe2001,Ansmann2009,Weinfurter2012}. Very recently it has been closed in optical setups \cite{Christensen2013,Giustina2013}, based on polarisation measurements of entangled photons distributed from a spontaneous parametric down-conversion (SPDC) source. These optical implementations represent an important achievement as they enable much higher rates of genuine random bits per time unit.

Given these experimental achievements, the natural question that arises is how to generate this genuine randomness efficiently. What is the maximal amount of randomness that a given physical implementation allows for? And most importantly, how should the relevant physical parameters of the setup be tuned to provide such an optimal amount? Here we answer these questions for the case of optical implementations based on SPDC, for which a thorough physical characterization has been recently presented in \cite{Vivoli2014}.

We start out by constructing a general framework and methods for optimal randomness certification in Bell experiments. The idea is to keep as much information as possible by avoiding any sort of binning of outcomes, then to use the methods recently introduced in \cite{Pironio2014} to estimate randomness by constructing a device-independent guessing probability optimized over all possible Bell inequalities, and finally to optimize the latter quantity over all the tunable physical parameters of the experiment. We then narrow our focus to entirely optical polarisation-based implementations (\textit{e.g.} \cite{Christensen2013,Giustina2013}). We first characterize the realistic parameters of such Bell setups and then apply our methods to determine optimal amounts of global and local randomness under realistic conditions. We provide interesting bounds on the experimental parameters -- some of them counter-intuitive and perhaps unexpected -- and certify up to four times more randomness than what a standard analysis, based on a binning of the outcomes and on the CHSH inequality \cite{CHSH}, can achieve \cite{Pironio2010}.

\section{Methods}
\label{sec.methods}

Here we describe methods that allow for optimal device-independent randomness certification. The general idea consists of three steps which are given in Box~1.  Since we do not make any physical characterization of the source or the devices, the results are kept general and can be applied to any bipartite Bell experiment free of the detection loophole (\textit{cf.} \cite{Rowe2001,Ansmann2009,Weinfurter2012,Christensen2013,Giustina2013}). 
\
\\
\
\\
\

\begin{footnotesize}
\textbf{Box 1.}  General directions for optimal randomness certification.

\end{footnotesize}

\fbox{
  \parbox[c][9em][c]{0.77\textwidth}{
\noindent 1. Estimate the most general behaviour $\textbf{p}$, without any binning. (\textit{Subsections \ref{sec.scenario} and \ref{sec.boundGp}})

\noindent 2. Construct $G_{\textbf{p}}$, the device-independent guessing probability optimized over all possible Bell inequalities. (\textit{Subsection \ref{sec.boundGp}})

\noindent 3. Optimize $G_{\textbf{p}}$ over the parameters  $\mathcal{P}$ that can be adjusted in the experimental setup.  (\textit{Subsection \ref{sec.expparam} and Section \ref{sec.allopt}})
  }
}
\\

\subsection{Scenario}
\label{sec.scenario}

To begin, we recall the device-independent scenario \cite{Pironio2010,Pironio2014,acin2007}. Two parties, Alice and Bob, are located in two secure laboratories from which no unwanted classical information can leak out. At each round of the experiment, they receive a quantum state $\rho_{\text{AB}}$ from a source S and perform on it one out of $m_A$ ($m_B$) possible measurements $x=0,1,..,m_A-1$ ($y=0,1,...,m_B-1$)  and retrieve one out of $o_A$ ($o_B$) possible outcomes $a=0,1,...,o_A-1$ ($b=0,1,..,o_B-1$). We make no other assumption on $\rho_{\text{AB}}$ other than the fact that it is a quantum state. In fact, $\rho_{\text{AB}}$ could have any dimension, and could even be correlated with another quantum system in the possession of a malicious eavesdropper Eve \footnote{We consider that Eve is limited by the laws of quantum mechanics. We also assume that the behaviour of the boxes is independent and identically distributed from one round to another, though, interestingly, the bound \eref{eq.optGp} has been proved secure under less demanding assumptions, (see \cite{Pironio2012}).}, such that $\rho_{\text{AB}} = \text{Tr}_{\text{E}} \rho_{\text{ABE}}$. 

Moreover, Alice and Bob do not trust the devices they use to measure $\rho_{\text{AB}}$. These devices can be thought of as measurements characterized by positive operator-valued measures (POVMs) with elements $\{M_{a|x}\}$ and $\{M_{b|y}\}$ acting on $\rho_{\text{AB}}$. Their probabilistic behaviour is given by Born's rule, 
\begin{equation}
\label{eq.behavdef}
p(ab|xy)=\text{Tr}[\rho_{\text{AB}}\ M_{a|x} \otimes M_{b|y}].
\end{equation}
There are a total of $m_Am_Bo_Ao_B$ such probabilities,  which can be seen as the components of a vector $\textbf{p}=\{ p(ab|xy)\} \in \mathbb{R}^{m_Am_Bo_Ao_B}$. We call $\textbf{p}$ the \textit{behaviour} associated with the \textit{quantum realization} \textit{Q} defined by the state $\rho_{\text{AB}}$ and the measurements with elements $\{M_{a|x}\}$ and $\{M_{b|y}\}$.

\subsection{Bounding the device-independent guessing probability}
\label{sec.boundGp}

The optimal amount of randomness that Alice and Bob can certify from an observed quantum behaviour $ \textbf{p} $ is measured here by the min-entropy of the \textit{device-independent guessing probability} $G_{\textbf{p}}$ \cite{Koenig2009}, \textit{i.e.} $h=-\log_{2}(G_{\textbf{p}})$. To estimate $G_{\textbf{p}}$, consider that for some round of the experiment Alice and Bob have chosen and performed some measurements $x=x^*$ and $y=y^*$  on $\rho_{\text{AB}}$. Without loss of generality any strategy $z$ of Eve can be seen as a POVM measurement with $o_Ao_B$ elements $\{M_{e|z}\}$ that she applies on her reduced state $\rho_{\text{E}}=\text{Tr}_{\text{E}} \rho_{\text{ABE}}$. Whenever she obtains the output $e=(a^*,b^*)$ she then guesses that Alice's (Bob's) outcome was $a^*$ ($b^*$). It can be shown that $G_{\textbf{p}}$, the average probability that Eve correctly guesses the output of Alice and Bob boxes using an optimal strategy, is the solution to the following conic linear program \cite{Pironio2014,Bancal2014}:
\begin{equation}
\label{eq.linprogGp}
 \left\{ \begin{array}{l}
G_{\textbf{p}}(x^*,y^*)\ =\ \ \underset{\{\textbf{p}^e\}}{\max} \ \ \ \ \ \sum\limits_{\substack{e}} p^e(e|x^*,y^*)
\\
\\ \text{s.t.}\ \sum\limits_{\substack{e}} \textbf{p}^{e} = \textbf{p}\ \ \text{and}\ \ \textbf{p}^{e} \in \widetilde{Q},\ \ \forall\ e=0,..,(o_A-1)(o_B-1).
\end{array} \right.
\end{equation}
Each $ \textbf{p}^{e} $ is an unnormalized behaviour ``prepared'' for Alice and Bob and conditioned on the outcome $e$ of the measurement with POVM elements  $\{M_{e|z}\}$ performed by Eve. Hence, the probability that $ \textbf{p}^{e} $ is prepared is the probability that Eve obtains the corresponding outcome $e$, \textit{i.e.} $p(e|z)=\text{Tr}[\rho_{\text{E}}\ M_{e|z}]$.  To be precise, $\textbf{p}^e=\{p^e(a,b|x,y)\}\in  \mathbb{R}^{m_Am_Bo_Ao_B}$, and $\widetilde{Q}$ is the set of all such unnormalized quantum behaviours. The first constraint in the program translates the fact that the behaviours $ \textbf{p}^{e} $ should on average reproduce Alice and Bob's observed behaviour $ \textbf{p} $. The second constraint demands that every behaviour should be quantum \footnote{A behaviour $\textbf{p}$ is said to be quantum whenever there exists a realization \textit{Q} (i.e. a quantum state + measurements) which reproduces $\textbf{p}$ through Born's rule \eref{eq.behavdef}.}. The program maximizes the success of Eve's strategy over all possible  $\{ \textbf{p}^{e}\ | e=0,...,(o_A-1)(o_B-1)\}$ decompositions.

The program presented in \eref{eq.linprogGp} is in general intractable due tu the lack of a precise characterization of $\widetilde{Q}$, but semi-definite programming (SDP) relaxations similar to the ones presented in \cite{Navascues2007} can be used tu put bounds on $G_{\textbf{p}}$. One then defines a convergent hierarchy of convex sets having a precise characterization and being such that $\widetilde{Q}_1 \supseteq \widetilde{Q}_2 \supseteq ... \supseteq  \widetilde{Q} $ \cite{Pironio2014,Navascues2007}. This hierarchy approximates the quantum set $\widetilde{Q}$̃ from the outside, and thus one can relax the difficulty of the problem (to the order $k$) by replacing $\widetilde{Q}$ in \eref{eq.linprogGp} by $\widetilde{Q}_k$. The solution $G^k_{\textbf{p}}$ of the $k$-th SDP program sets an upper bound on the guessing probability $G_{\textbf{p}}$, which in turn sets a lower bound $h^k=-\log_2(G^k_{\textbf{p}})$ on the number $h$ of \textit{global} random bits that are certified from $\textbf{p}$ and from the measurements $(x^*,y^*)$.

It is worth mentioning that the methods presented so far can be adapted straightforwardly for \textit{local} randomness evaluation. In this case, the situation is considered from Alice's perspective, for example, and a program equivalent to \eref{eq.linprogGp} is derived to obtain the local guessing probability $G_{\textbf{p}}(x^*)$. Computationally speaking, local randomness is appealing as the number of POVM elements of Eve's strategies gets reduced from $o_Ao_B$ to $o_A$. 

To conclude this section notice that the optimal Bell inequality which yields $G^k_{\textbf{p}}$ can be accessed from the dual formulation of \eref{eq.linprogGp}. The advantage with respect to previous methods (which assess the problem via a fixed Bell inequality, \textit{e.g.} \cite{Pironio2010}) has been found to be significant in both \cite{Pironio2014} and \cite{Bancal2014,Bancal2014b,Bancal2014c}.

\subsection{Keeping as much data as possible}
\label{sec.keepdata}

In subsection \ref{sec.boundGp} we discussed how to quantify the maximal amount of randomness available for Alice and Bob from an observed behaviour $ \textbf{p}$. Still, there are several degrees of freedom in  $ \textbf{p}$ that can be further optimized to provide even more randomness. More precisely, tailoring these degrees of freedom always leads to different behaviours, which in turn yields different --and hopefully higher-- amounts of randomness. We can distinguish two types of such degrees of freedom; those that require adjustments in the experimental setup (\textit{e.g.} increasing the efficiency of the detectors), and those which do not. Here we will deal with the latter, and leave the former for subsection \ref{sec.expparam}. 

In particular, the numbers of outcomes $o_A$ and $o_B$ can be adjusted without much experimental effort. All Bell experiments so far, which have managed to close the detection loophole, have relied violation of the CHSH inequality \cite{CHSH} (or similar ones \cite{Eberhard93}). This assumes the local observation of two outcomes per party. However, in addition to the two good outcomes, loss and imperfections lead to events where no detector clicks, resulting in a third outcome per party; this means that a local binning process was applied in all these experiments to reduce the size of the original behaviour to two outcomes.

It is intuitive to expect that more randomness can be certified when binning strategies are avoided; any binning strategy represents a loss of potentially useful information. Still, it could be the case that the amount of certifiable randomness would not get diminished for some particular binning. Our results in section 4 show that this is not the case in general. In fact, In \ref{app.CHSHineff} we explicitly show how any binning strategy applied to CHSH correlations with inefficient detectors will systematically decrease the amount of certifiable randomness. Hence, to certify optimal amounts of randomness, Alice and Bob must ensure that the number of outcomes $o_A$ and $o_B$ is kept as high as possible.
 
\subsection{Taking experimental parameters into account}
\label{sec.expparam}

The observed quantum behaviour $\textbf{p}$ possesses physical degrees of freedom that can be adjusted in the experimental setup to produce higher amounts of randomness. The solution of \eref{eq.linprogGp} can be minimized over all the possible realistic values that such parameters (which we label $\mathcal{P}$) can take. In this way, the optimal amount of randomness that can be certified to the order $k$ is the solution of:
\begin{equation}
\label{eq.optGp}
 \left\{ \begin{array}{l}
G^k(x^*,y^*)\ =\ \min\limits_{\substack{\mathcal{P}}} \ \ G^k_{\textbf{p}}(x^*,y^*)
\\
\\ \text{s.t.}\ \ G^k_{\textbf{p}}(x^*,y^*) \text{ solves the $k$th SDP of (2)}.
\end{array} \right.
\end{equation}
In particular, notice that this program optimizes $G^k_{\textbf{p}}(x^*,y^*)$ over the number of measurements $m_A$ and $m_B$, which are implicit quantities in $\mathcal{P}$ (see also section \ref{sec.constPpG}). 
\
\\
\

\section{Realistic optical implementations }
\label{sec.allopt}

\begin{figure}[h]
\centering
\includegraphics[width = 13cm]{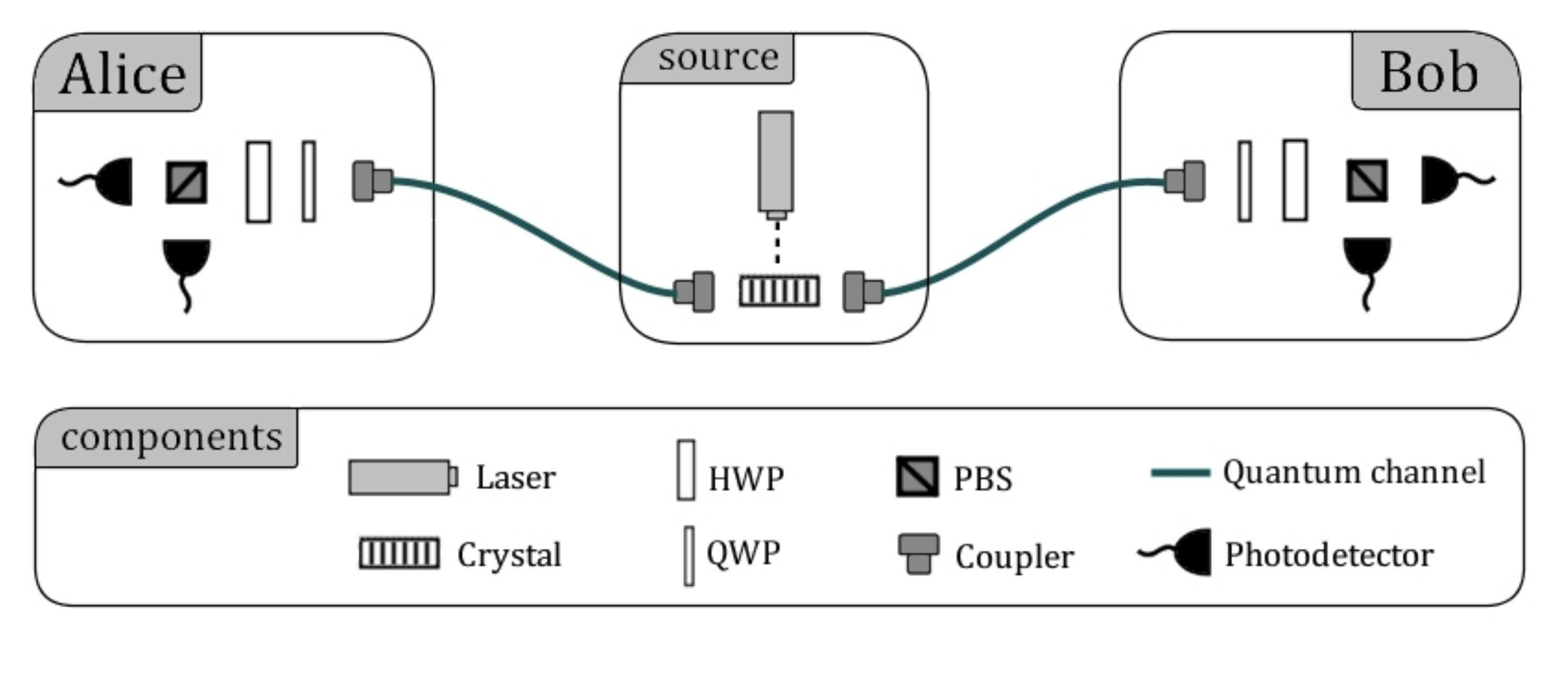}
\caption{Experimental setup for optical Bell experiments based on SPDC.}
\label{fig.setup}
\end{figure}

The methods presented above are general and can be adjusted to any bipartite Bell experiment. We focus and describe in the following the architecture of optical implementations based on polarisation measurements of entangled photons distributed from an SPDC source (see \figref{fig.setup}), which was thoroughly analysed in \cite{Vivoli2014}. The source is characterized by three adjustable quantities: two squeezing parameters $g_1$ and $g_2$  and a total number of modes $N$ onto which the photons may be distributed. Each mode locally splits into two orthogonal polarisations. In terms of bosonic creation operators, the unnormalized state produced by S is given by \cite{Vivoli2014}:
\begin{equation}
\label{eq.spdcState}
\prod\limits_{k=1}^{N}\ \exp\left[\tanh(g_1)a_k^{\dagger}b_{k\perp}^{\dagger} - \tanh(g_2)a_{k\perp}^{\dagger}b_{k}^{\dagger} \right]\ \ket{0},
\end{equation}
were $\ket{0}$ is the vacuum state associated to the $4N$ bosonic operators $a_{1}^{\dagger},...,a_{N\perp}^{\dagger},b_{1}^{\dagger},...,b_{N\perp}^{\dagger}$, and the $a$-modes ($b$-modes) are distributed to Alice (Bob).

All the different types of losses including detectors inefficiencies are modelled, without loss of generality, by two beam-splitters (not shown in \figref{fig.setup})  placed at any point between the users and the source. The transmittance $\eta$ of these beam-splitters is the overall detection efficiency of the experiment. 

The measurements are performed with polarizing beam-splitters (PBS) and half-wave plates (HWP) and quarter-wave plates (QWP) which allow splitting the orthogonal modes along arbitrary directions \cite{Christensen2013,Giustina2013,Vivoli2014}. Each measurement $u$ is fully characterized by two angles $(\theta_u,\phi_u)$ defining a projection in the Bloch sphere. Each of the parties holds two detectors, which do not resolve photon number. Hence, for each detector only the outcomes ``0=No click" and ``1=Click" can be distinguished, and the maximal number of local outcomes (without binning) is $o_A=o_B=4$.

\section{Results}
\label{sec.results}

In this section we apply the methods presented in section~\ref{sec.methods} to the optical setup described in section~\ref{sec.allopt}. 

\subsection{Constructing $\mathcal{P}$, \textbf{p} and $G$}
\label{sec.constPpG}

Considering that Alice and Bob respectively perform $m_A$ and $m_B$ measurements, the experiment is characterized by $4+2(m_A+m_B)$ physical parameters, which are: $N$, $g_1$, $g_2$, $\eta$, $\theta_1^A$, $\phi_1^A$, ... , $\theta_{m_B}^B$ and $\phi_{m_B}^B$. All of these parameters are adjustable within some range of realistic values, except $\eta$ which, as discussed above, represents the main restriction for an optical implementation. Hence, the adjustable parameters read:
\begin{equation}
\label{eq.Pdef}
\mathcal{P} = (N,g_1,g_2,\theta_1^A,\phi_1^A,...,\theta_{n_B}^B,\phi_{n_B}^B).
\end{equation} 

The analytic expression of $\textbf{p}$ as a function of $\mathcal{P}$ and $\eta$ is at first only computed for the first measurements of Alice and Bob, $(\theta_1^A,\phi_1^A)$ and $(\theta_1^B,\phi_1^B)$. In this case $\mathcal{P}$ consists of seven parameters, \textit{i.e.} $\mathcal{P} = (N,g_1,g_2,\theta_1^A,\phi_1^A,\theta_1^B,\phi_1^B)$. Since the number of outcomes are kept as high as possible ($o_A=o_B=4$), this expression is obtained by solving a linear system of $4\times 4=16$ equations; $15$ of these equations correspond to the ``no-click'' probabilities of all the detectors, which can be found in the supplementary material of \cite{Vivoli2014}. The remaining equation is a normalization condition. 

Next, this expression (obtained only for the first measurements) is generalized for arbitrary $(m_A,m_B)$. One only needs to concatenate all the individual behaviours:
\begin{equation}
\label{eq.pStructure}
\fl \ \ \ \ \ \ \ \ \ \textbf{p}= \left\{\textbf{p}(N,g_1,g_2,\eta,\theta_i^A,\phi_i^A,\theta_j^B,\phi_j^B)\ \ |\ \ 1\leq i \leq m_A\ \ \text{and}\ \ 1\leq j \leq m_B\right\}.
\end{equation}
In particular, all the individual behaviours have the same analytical structure as the behaviour obtained for the first measurements, and hence one only needs to substitute $\theta_1^{A}\leftarrow \theta_i^{A}$, $\theta_1^{B}\leftarrow \theta_j^{B}$, $\phi_1^{A}\leftarrow \phi_i^{A}$ and $\phi_1^{B}\leftarrow \phi_j^{B}$ for each $i$ and $j$ in \eref{eq.pStructure}. This yields the desired $m_A m_B o_A o_B$-sized quantum behaviour (see subsection \ref{sec.scenario}).

Finally, it is necessary to set realistic limits on $\mathcal{P}$; otherwise, the minimization in \eref{eq.optGp} is unbounded. We let $1\leq N \leq 100$, $-1/2\leq g_1,g_2 \leq 1/2$ (corresponding to about $4.3$ dB of squeezing) and we let all the measurement angles vary in a $2\pi$-length interval.

\subsection{Optimal randomness for $m_A=m_B=2$}
\label{sec.optrndn2}

Optimal randomness is retrieved from \eref{eq.optGp} upon optimization of all adjustable parameters, which include the number of measurements in the experiment. Optimizing $G^k$ over $m_A$ and $m_B$ is of particular relevance for the setup that we consider as distinct rotation directions of the incoming modes can be achieved by adjusting the HWP and QWP, \textit{i.e.} without the need of further experimental resources. Still, to illustrate the performance of our methods we consider here the simplest case $m_A=m_B=2$.

\begin{figure}[t]
\centering
\includegraphics[width = 12cm]{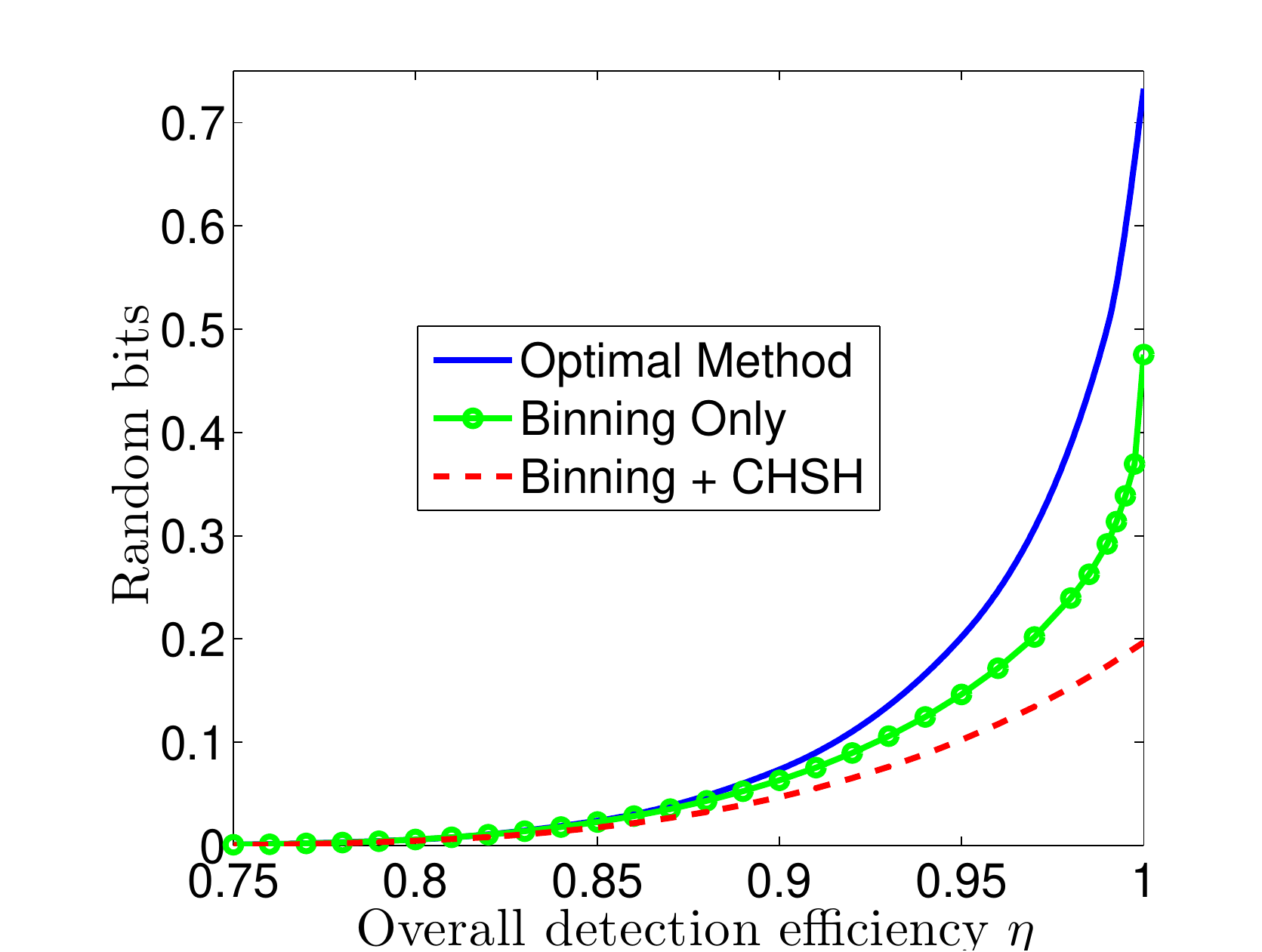}
\caption{Global randomness for the case $m_A=m_B=2$. For the three curves, the parameters $\mathcal{P}$ are optimized at each point, as explained in subsection \ref{sec.expparam}. The solid curves are the min-entropy of the solution of program \eref{eq.optGp} for $o_A=o_B=4$ (optimal) and for $o_A=o_B=2$ (binning). The dashed and dotted curves where obtained following the binning strategy presented in \cite{Vivoli2014}.}
\label{fig.Gp22}
\end{figure}

\begin{figure}[t]
\centering
\includegraphics[width = 1\textwidth]{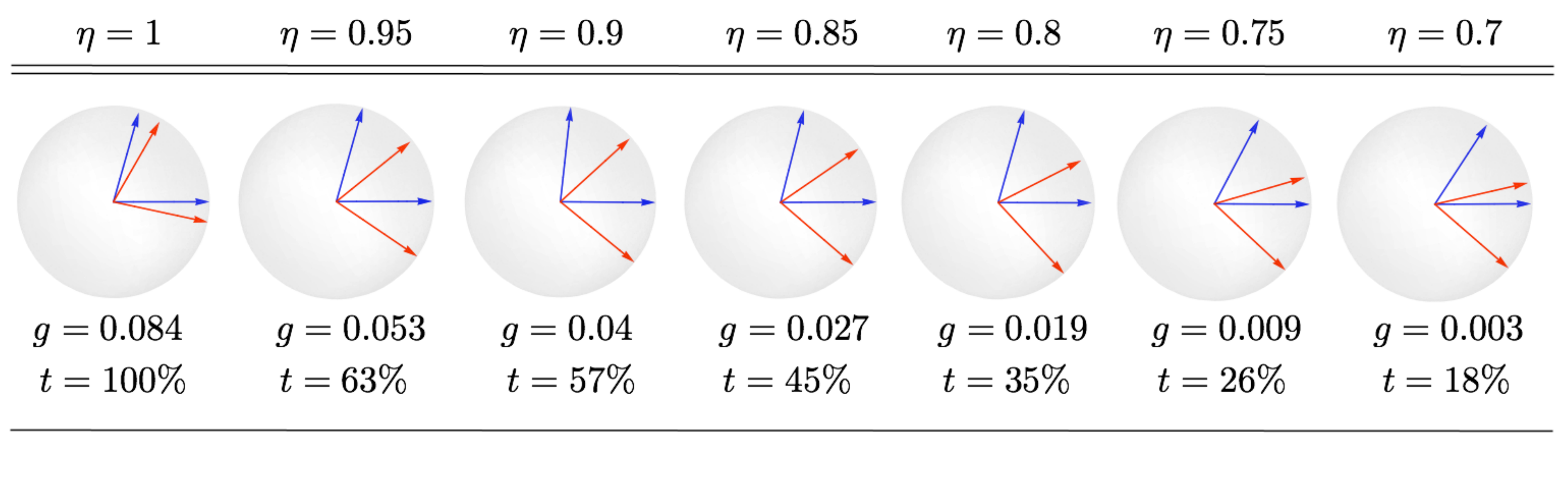}
\caption{Color online. Optimal parameters $\mathcal{P}$ for different values of $\eta$. $t$ is the ratio between $\tanh(g_1)$ and $\tanh(g_2)$, while $g=\max (g_1,g_2)$. $N$ always reaches $100$.  Blue (Red): optimal measurements for Alice (Bob) in the Bloch sphere representation. All these quantities were obtained after solving program \eref{eq.optGp}.}
\label{fig.optP}
\end{figure}

We find \footnote{All our results were obtained at the order $k=1+AB$. This corresponds to an intermediate stage $\widetilde{Q}_1 \supseteq \widetilde{Q}_{1+AB} \supseteq \widetilde{Q}_2$; see \cite{Navascues2007} for details.} that whenever the parties are restricted to $o_{\text{bin}}=2$ outcomes, more global randomness is certified when no specific Bell inequality is considered. This was to be expected following subsection \ref{sec.boundGp} and the line of research of \cite{Pironio2014,Bancal2014,Bancal2014b} (see dashed and dotted curves in \figref{fig.Gp22}). However, we improve considerably this expected result by suppressing the binning of the outcomes and letting $o=4$, as we explained in subsection \ref{sec.keepdata} (solid curve in \figref{fig.Gp22}). For $\eta=1$ our methods certify $0.74$ bits of global randomness per source use, four times more than the $0.19$ bits that are certified from the CHSH inequality (we provide the Bell inequality that certifies this improvement in \ref{app.Bellineq}). 
The numerical values of the optimal parameters $\mathcal{P}$ are given in \figref{fig.optP} for several values of $\eta$. Intuitively, the ratio $t=\tanh(g_1)/\tanh(g_2)$  quantifies the degree of entanglement of the source, as \eref{eq.spdcState} shows.  For $\eta=1$ optimal randomness is obtained from a ``maximally entangled'' state, \textit{i.e.} $t=100\%$, but as $\eta$ decreases $t$ also decreases. This was to be expected for the lower values of $\eta$, where nonlocality can only be certified with non-maximally entangled states \cite{Eberhard93}. Interestingly,  for $\eta\approx1$  the optimal measurements are not similar to the ones that intuitively maximize the violation of the CHSH inequality on two maximally entangled qubits (\textit{e.g.} they are not mutually unbiased); see \ref{app.Bellineq} for the exact expressions. That is, the optimal measurements for optimal randomness certification are not the same as those maximizing the CHSH violation.

The number of modes attains the maximal value that we allow ($N=100$) whenever $\eta$ is greater than $2\sqrt{2}-2$. For $\eta$  smaller than this value, the single mode case $N=1$ is sufficient to obtain maximal randomness; this fact was noticed in  \cite{Vivoli2014} for the maximization of the CHSH inequality violation. Finally, we have found that the improvement obtained when increasing the number of modes beyond $\approx 25$  is very small.

\subsection{Optimal randomness with more than two measurements}

Our next goal is to see whether deploying more measurements yields an improvement in the number of random bits. In the previous subsection we considered the case $m_A=m_B=2$; however, by adjusting the HWP and QWP located in front of their PBS, Alice and Bob can measure their incoming subsystem along any arbitrary polarisation direction of the Bloch sphere. These adjustments can thus be obtained with relatively low experimental cost, the main drawback being a non-negligible increase in the amount of statistical data (the size of the observed behavior $\textbf{p}$ increases with $m_Am_B$).

Our results in \tabref{tab.rndbits} show that more measurements certify more randomness, even in scenarios for which a binning strategy had to be considered and $\mathcal{P}$ could not be fully optimized due to computational limitations. The time required to solve \eref{eq.optGp} becomes large as the number of measurements increases, since the total number of SDP variables describing the behaviours $\textbf{p}^e$ in \eref{eq.linprogGp} increases as $(m_Am_B)^2$. The increase is less dramatic when \textit{local} randomness is certified \textit{e.g.} from Alice's perspective, as there are only $o_A$ (instead of $o_Ao_B$) SDP matrices in \eref{eq.linprogGp} for each choice of $\mathcal{P}$.
\begin{table}[h!]
\renewcommand{\tabcolsep}{0.2cm}
\centering
 \begin{tabular}{l*{2}{c}*{4}{c}}
     SCENARIO          & $(2,2)$             & $(3,2)$       & $(3,3)$    & $(4,3)$  & $(4,4)$      & $(5,5)$                \\  [1ex]
     \hline \hline \\ [0.03 cm]
     Total SDP variables   & $1348$ & $3340$                                  & $8392$   & $15748$ & $29620$             & $\sim 10^5$                                      \\  [1ex]
     Local random bits  & $0.454$ & $0.459$                                  & $0.519^*$ & $0.523^*$ & $0.557^*$              & N/A                                   \\  [1ex]
     \hline
 \end{tabular}
\caption{Local randomness certified for different scenarios for $\eta=1$. The scenario specifies the couple $(m_A,m_B)$. The * symbol is used when full optimization was not possible, and instead: \textit{(i)} the optimization was only carried over the number of modes, with $g_1=g_2=0.1$; \textit{(ii)} the measurements were inspired from the chained inequality \cite{Braunstein90} and \textit{(iii)} we considered $3$ outcomes per party by locally binning the ``no click-no click'' and the ``click-click'' outcomes.}
\label{tab.rndbits}
\end{table}

In particular, with four measurements per party we certify $0.557$ local random bits. This is $3$ times more than the amount that is certified from the CHSH inequality ($\approx 0.17$ bits) under the same considerations.
  
\subsection{Experiments with only one detector per side}

The setup depicted in \figref{fig.setup} has been hitherto central in our analysis as it captures the general architecture for Bell experiments with entangled photons. Unfortunately, state-of-the-art superconducting detectors, \textit{i.e.} those which achieve detection efficiencies above $70\%$ and thus enable a true Bell violation without post-selection, represent an extremely high experimental cost nowadays.

This situation can be alleviated (the cost can be reduced by half) by realizing that a Bell test can still be carried on with the use of only one detector on each arm of the experiment \cite{Christensen2013,Giustina2013}. Given the techniques that we have shown so far, it is interesting to see how the optimal amount of randomness is affected. For a fixed overall detection efficiency $\eta$, how does the optimal amount of randomness that can be certified in an experiment with only one detector compare to the optimal amount of randomness that can be certified with two detectors?

The statistics of an experiment with only one detector are straightforwardly obtained from the statistics of an experiment with two detectors (those which we presented in \ref{sec.constPpG}). As discussed in \ref{sec.allopt} the possible local outcomes of an experiment with two detectors are 00, 01, 10 and 11 where the first (second) number labels the outcome of the first (second) detector``0=No click" and ``1=Click''. Then, applying the local binning $\mathcal{B}_{\text{1Det}}=\{00\rightarrow0',\ 01\rightarrow0',\ 10\rightarrow1',11\rightarrow1'\}$ on Alice and Bob's sides yields the statistics of the experiment without the second detector.

We observe that for $\eta \lesssim 0.8 $ no disadvantage occurs if the second detector is removed: the optimal amount of local and global randomness than can be certified in both cases is $\sim 6\times 10^{-4}$ bits. On the other hand, as $\eta$ becomes close to $1$ removing a detector negatively affects the optimal amount of randomness: for $\eta=1$ the optimal amount of local (global) random bits certified with two detectors is $\approx 0.45$ ($\approx 0.73$) bits, while with only one detector the optimal amount is $\approx 0.31$  ($\approx 0.34$) bits.

\section{Discussion}
Summarizing, in the present article we have explicitly shown the benefits of optimizing randomness in a Bell experiment over all possible inequalities, and the negative consequences that occur when information is lost through a binning of the resulting outcomes.  We carefully analysed and characterized optical setups based on SPDC and certified up to four times more randomness when all of the physical parameters were optimized. 

To put it in a nutshell, here are the important facts to be aware of in order to retrieve optimal amounts of randomness from an optical Bell implementation based on SPDC \textit{(and their experimental cost)}:
\
\\
\

\textbf{\textit{1.}} Keep the whole statistics and avoid binning the outcomes. \textit{(no cost).}

\textbf{\textit{2.}} Use as many polarisation measurements as possible. \textit{(small cost).}

\textbf{\textit{3.}} Use many modes to distribute the entangled photons. \textit{(high cost in principle,}
 
\indent \textit{but keep in mind that more than $\approx 25$ modes will provide little improvement).}

\textbf{\textit{4.}} For $\eta \approx 1$, the optimal measurements for randomness extraction are not

\indent the ones that maximize the violation of the CHSH inequality. \textit{(no cost).}

\textbf{\textit{5.}} For $\eta \lesssim 0.8$ it is enough to use a single mode to distribute entanglement

\indent and use a single detector per side. \textit{(no cost).}
\
\\
\

 We hope that this work will be useful for the future development of Bell-type randomness generation experiments. 

\section*{Acknowledgments}

We thank M Hoban and S Pironio for interesting discussions and for the proof presented in \ref{app.CHSHineff}. We also thank N Sangouard for sharing with us the exact expressions of the no-click probabilities discussed in subsection \ref{sec.constPpG}. The SDP calculations were performed using the code QMBOUND written by JD Bancal. This work was supported by  by the EU projects QITBOX and SIQS and the John Templeton Foundation. JB was supported by the Swiss National Science Foundation (QSIT director's reserve) and SEFRI (COST action MP1006), DC by the Beatriu de Pin\'os fellowship (BP-DGR 2013), AM by the Mexican CONACYT graduate fellowship porgram, and PS by the Marie Curie COFUND action through the ICFOnest program. 
\
\\
\
\\
\
\\
\

\appendix

\section{CHSH correlations with inefficient detection} 
\label{app.CHSHineff}

Here we show that it is always advantageous to keep the ``no-click'' outcome in a CHSH test with inefficient detection. Assume that at each round of the experiment Alice and Bob receive a perfect singlet, \textit{i.e.} a maximally entangled state of two qubits, on which with equal probability Alice measures $\sigma_z$ and $\sigma_x$, while Bob measures $(\sigma_z + \sigma_x)/2 $ and $(\sigma_z - \sigma_x)/2 $. If the measurement processes have non-unit $\eta$ efficiency, the possible outcomes that the users observe are $0$, $1$ and $2$ (here the outcome $2$ labels the no-click outcome). Under the assumption that losses occur independently, the observed quantum behaviour can be written as 

\begin{equation}
\label{eq.peta}
\textbf{p}_{\eta}\ =\ \ \left. \begin{array}{lll|l}
\eta^2c & \eta^2s &  \frac{\eta(1-\eta)}{2} & ~
\\
\eta^2s & \eta^2c &  \frac{\eta(1-\eta)}{2} & \vdots
\\
\frac{\eta(1-\eta)}{2} & \frac{\eta(1-\eta)}{2} &  (1-\eta)^2 & ~
\\ \hline
 ~ & \cdots &  ~ & \ddots
\end{array} \right.
\end{equation}

\noindent with $c,s = (2\pm \sqrt{2})/8$. In this expression each of the 4 blocks describes the joint probability $P(a,b|x,y)$ for a choice of measurements of Alice and Bob. The first block corresponds to $(x=0,y=0)$ and so on. Blocks 2 and 3 are equal to block 1, while a swap between $c$ and $s$ transforms block 1 into block 4. For each choice of $x$, any \textit{physical} binning is a deterministic map from the outcomes $\{(a=0,a=1,a=2)\}$  into the binned outcomes $(a'=0,a'=1)$, and the same applies to each choice of $y$ with $b$. Up to local relabelings, there are only three relevant binning strategies (three ways to bin a local trit to a bit) which are, with a slight abuse of notation, $\mathcal{B}=\{0\rightarrow0',\ 1\rightarrow1',\ 2\rightarrow0'\}$, $\mathcal{B'}=\{0\rightarrow0',\ 1\rightarrow1',\ 2\rightarrow1'\}$ and $\mathcal{B''}=\{0\rightarrow0',\ 1\rightarrow0',\ 2\rightarrow1'\}$. However, $\mathcal{B''}$ is not relevant as it erases all non-local data. Hence we are left with two local binning strategies which in turn generate four possible quantum behaviours for Alice and Bob:
\begin{equation}
\label{eq.ABbehavs1}
\fl \textbf{p}_{\mathcal{B}\mathcal{B}}=\left. \begin{array}{cc|c}
\eta^2c+1-\eta & \eta^2s+\frac{\eta(1-\eta)}{2} & ~
\\
\eta^2s+\frac{\eta(1-\eta)}{2} & \eta^2c &  \vdots
\\ \hline
 ~ & \cdots &  \ddots
\end{array} \right.;
\ \textbf{p}_{\mathcal{B'}\mathcal{B'}}=\left. \begin{array}{cc|c}
\eta^2c & \eta^2s+\frac{\eta(1-\eta)}{2} & ~
\\
\eta^2s+\frac{\eta(1-\eta)}{2} & \eta^2c+1-\eta &  \vdots
\\ \hline
 ~ & \cdots &  \ddots
\end{array} \right.
\end{equation}
and
\begin{equation}
\label{eq.ABbehavs2}
\fl \textbf{p}_{\mathcal{B}\mathcal{B'}}=\left. \begin{array}{cc|c}
\eta^2c+\frac{\eta(1-\eta)}{2} & \eta^2s+1-\eta & ~
\\
\eta^2s & \eta^2c+\frac{\eta(1-\eta)}{2} & \vdots
\\ \hline
 ~ & \cdots &  \ddots
\end{array} \right.;
\ \textbf{p}_{\mathcal{B'}\mathcal{B}}=\left. \begin{array}{cc|c}
\eta^2c+\frac{\eta(1-\eta)}{2} & \eta^2s & ~
\\
\eta^2s+1-\eta & \eta^2c+\frac{\eta(1-\eta)}{2} &  \vdots
\\ \hline
 ~ & \cdots &  \ddots
\end{array} \right. .
\end{equation}

\begin{figure}[h!]
\begin{center}
\includegraphics[width = 13cm]{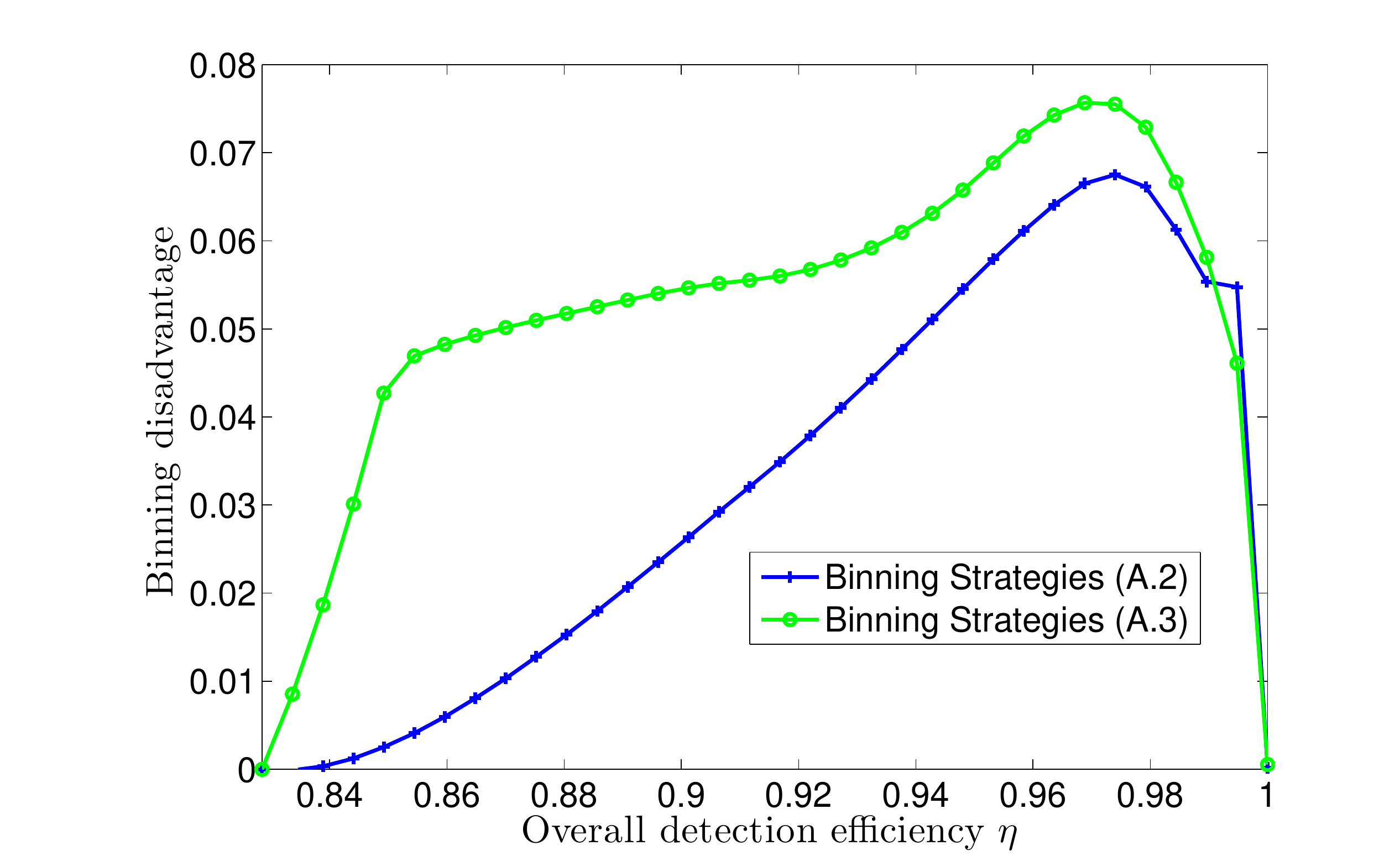}
\end{center}
\caption{The binning disadvantage is the difference in the number of bits that are certified from either $\textbf{p}_{\mathcal{B}\mathcal{B}}$ or $\textbf{p}_{\mathcal{B}\mathcal{B'}}$ with respect to $\textbf{p}_{\eta}$.}
\label{fig.bindisadv}
\end{figure}

Notice from \eref{eq.ABbehavs1} that whenever Alice and Bob apply the same binning strategy the two resulting probability distributions have the same values up to a permutation of the elements. The same occurs in \eref{eq.ABbehavs2} whenever they apply a different binning. It is therefore sufficient to evaluate the optimal randomness available from $\textbf{p}_{\mathcal{B}\mathcal{B}}$ and from  $\textbf{p}_{\mathcal{B}\mathcal{B'}}$, for example. In \figref{fig.bindisadv} we plot the percentage by which the guessing probability for these quantum behaviours is increased with respect to the guessing probability obtained from $\textbf{p}_{\eta}$. We find that for any $2\sqrt2-2<\eta<1$ it is always advantageous to keep the no-click outcome.

\section{Bell Inequality and relevant parameters expressions} 
\label{app.Bellineq}

As explained in the main text, the dual formulation of \eref{eq.linprogGp} yields the expression of the Bell inequality that certifies the optimal amount randomness \cite{Pironio2014}. It is therefore possible to retrieve the Bell inequality associated to the optimal parameters. One first solves the program \eref{eq.optGp} for $\eta$ fixed; this yields some optimal parameters $\mathcal{P}=\mathcal{P}^*$. Then one comes back to solve the dual program of \eref{eq.linprogGp} using as input $\textbf{p}(\mathcal{P}^*)$. In the Collins-Gisin parametrization, the $7\times7$ Bell inequality which certifies $0.74$ bits of global randomness (see subsection \ref{sec.optrndn2}) is:
\begin{equation}
\fl \ \ I_{\eta=1}^{1+AB}\ \ =\ \ \left. \begin{array}{c|ccc|ccc}
1 & 8.02 & 8.18  & 8.18 & 8.11 & 12.38 & 12.37
\\ \hline
8.02 & -8.07 & 8.13 & 8.13 & -8.11 & 7.11 & 7.11
\\ 
8.18 & 8.13 & -2.80 & 6.68 & 7.53 & 19.63 & -20.54
\\ 
8.18 & 8.13 &  6.68 & -2.80 & 7.53 & -20.54 & 19.64
\\ \hline
8.11 & -8.11 & 7.53 & 7.53 & -7.98 & 7.77 & 7.77
\\ 
12.37 & 7.11 & 19.64 & -20.54 & 7.77 & 3.92 & -6.71
\\ 
12.37 & 7.11 & -20.54 & 19.64 & 7.77 & -6.71 & 3.92
\end{array} \right.,
\end{equation}
and the optimal parameters which enable this realization are (\textit{cf.} \eref{eq.Pdef}):
\begin{equation}
\fl \mathcal{P} = (100,\ 0.084,\ 0.084,\ 2.088,\ 1.116,\ 1.473,\ 1.117,\ 1.36,\ 1.117,\ 1.976,\ 1.116).
\end{equation}

\section*{References}

\bibliographystyle{iopart-num}
\bibliography{biblio}

\end{document}